\documentstyle{article}
\renewcommand{\theequation}{\thesection.\arabic{equation}}
\newcommand\beq{\begin{equation}}
\newcommand\eeq{\end{equation}}
\newcommand{\eqref}[1]{(\ref{#1})}

\def\beqa{\begin{eqnarray}}
\def\eeqa{\end{eqnarray}}
\def\ba{\begin{array}}
\def\ea{\end{array}}

\def\la{\lambda}

\def\lt({\left(}
\def\rt){\right)}

\newcommand{\bl}[1]{\makebox[#1em]{}}
\begin{document}
\hfill{PDMI PREPRINT --- 17/1997}
\footnotetext[1]{E-mail: bogoliub@pdmi.ras.ru}
\footnotetext[2]{E-mail: izergin@pdmi.ras.ru} 
\footnotetext[3]{E-mail: kitanin@pdmi.ras.ru}  
\vspace{20mm}

\centerline{\Large {\bf Correlation functions}}
\centerline{\Large {\bf for a strongly correlated boson system.}}

\vspace {10mm}
\centerline{\large {\bf N.M. Bogoliubov\footnotemark[1], A.G. Izergin\footnotemark[2]
, N.A. Kitanine\footnotemark[3]}}{%

\vspace{5mm}

\centerline {\it St. Petesburg Department of V.A. Steklov Mathematical
Institute,} \centerline{\it Fontanka 27, St. Petersburg 191011, Russia}

\begin{abstract}
The correlation functions for a strongly correlated exactly solvable
one-dimensional boson system on a finite chain as well as in the
thermodynamic limit are calculated explicitly. 
This system which we call the phase model 
is the strong coupling limit of the integrable $q$-boson hopping model. 
The results are presented as
determinants.
\end{abstract}

\section*{Introduction}

In this paper we investigate a one-dimensional strongly correlated boson 
lattice system 
 where the kinetic term depends on the occupation of
the adjacent lattice sites. It was introduced and solved by the quantum
inverse scattering method (QISM) in \cite{bbg,bb,bbgt,bbt}. The natural
dynamical variables for this model are the so-called $q$-bosons \cite{kul1}
closely related to the quantum algebra formalism \cite{kul2}. This model is
called the $q$-boson hopping model. It can be treated as a quantization \cite
{kul} of the classical Ablowitz-Ladik equation \cite{ablad} which is one of
the possible integrable lattice versions of the nonlinear Schr\"odinger
equation \cite{k}. In the scaling continuous limit the $q$-boson hopping
model becomes the Bose gas model.

We calculate correlation functions in the special case of the
lattice $q$-boson model corresponding to the infinite value of the coupling
constant. It appears that this special case is related to the phase
operators of the quantum nonlinear optics so we will call it the phase
model \cite{bik1}. Usually the infinite coupling limit
corresponds to the free fermion point (the most well-known example is the
Bose gas model). On the contrary the phase model cannot be regarded as a
free fermion model.

The description of the correlation functions for the models solved by means
of Bethe Ansatz is based on the representation for the correlators as the
Fredholm determinants of linear integral operators. Such representations
were obtained for the first time in \cite{l1,l2} for the simplest two point
equal-time correlators of one-dimensional impenetrable bosons. Later they
were generalized for the case of time-dependent correlators for the models
which are the free-fermion points of models solved by means of the quantum
inverse scattering method (the Bose gas in the infinite coupling limit \cite
{ks} and the isotropic XX0 Heisenberg chain \cite{cikt,cikt2}) and also for the
case of finite interaction (\cite{kor,kks} for the one-dimensional Bose gas
and \cite{efik} for XXZ chain). Such representations give an opportunity to
get classical integrable equations for correlators which can be used, in
particular, to calculate the long time and large distance asymptotics of
correlation functions \cite{iik1,iiks2,iik2,iikv}.

The main result of this paper is the determinant representations for
correlation functions. We consider the finite chain as well as the
thermodynamic limit at zero and finite temperature. The form factors for the
phase model were calculated in \cite{kit}. Some of the results in the
thermodynamic limit were reported in \cite{bik1}.

This paper is organized as follows. In the first Section the $q$-boson model
is defined and different limits are discussed. In Section 2 the Bethe Ansatz
solution for the $q$-boson hopping model is presented. In Section 3 we
discuss the solution of the phase model and its thermodynamics. In Section 4
the scalar products are calculated. In Section 5 we derive the determinant
representation for the simplest correlation function which is the ''darkness
formation probability''. In Section 6 we calculate the two-point
time-dependent correlation functions for the finite lattice while the
thermodynamic limit is obtained in the last Section. The derivation of the
form factors is presented in Appendix.

\section{Q-Boson hopping model}

\setcounter{equation}{0} The $q$-boson hopping model is defined by the
Hamiltonian 
\begin{equation}
H_q=-\frac 12\sum_{n=1}^M(B_n^{\dagger }B_{n+1}+B_nB_{n+1}^{\dagger }-2N_n),
\label{hamgen}
\end{equation}
with the periodic boundary conditions, $M+1\equiv 1$. Operators $B_n,B_n^{\dagger
},$ and $N_n=N_n^{\dagger }$ form the $q$-boson algebra: 
\begin{equation}
\lbrack N_i,B_j^{\dagger }]=B_i^{\dagger }\delta
_{ij}\,,\,[N_i,B_j]=-B_i\delta _{ij},  \label{qbos}
\end{equation}
\[
\lbrack B_i,B_j^{\dagger }]=q^{-2N_i}\delta _{ij}. 
\]
The c-number $q$ is taken to be $q=e^\gamma .$ We consider real $\gamma >0$.
This algebra was investigated in \cite{kul1} and the connection with the
quantum $SU_q(2)$ algebra was established in \cite{kul2}.

The Hamiltonian (\ref{hamgen})of the  
$q-$boson hopping model commutes with the
operator of total number of particles, 
\begin{equation}
\hat N=\sum_{n=1}^MN_n\,,\,[H_q,\hat N]=0\,.  \label{nh}
\end{equation}
The Hamiltonian with the chemical potential $\bar \mu $ 
\begin{equation}
H_h=H_q-\bar \mu \hat N,  \label{hamhim}
\end{equation}
which has the same eigenstates will be considered further.

The $q-$boson algebra (\ref{qbos}) possesses the representation in the local
Fock spaces formed by the $q$-boson normalized states 
\begin{equation}
B_j\mid 0\rangle _j=0,\,\,\,N_j\mid 0\rangle _j=0,
\label{frep}
\end{equation}
\[
B_j^{\dagger }\mid n_j\rangle _j=[n_j+1]^{\frac 12}\mid n_j+1\rangle
_j\,,\,\,\,\,B_j\mid n_j\rangle _j=[n_j]^{\frac 12}\mid n_j-1\rangle _j, 
\]
and the normalized states on the whole lattice are 
\begin{equation}
\mid 0\rangle =\prod_{j=1}^M\mid 0\rangle _j\,,  \label{qstate}
\end{equation}
\[
\mid n\rangle =\prod_{j=1}^M\mid n_j\rangle _j=\prod_{j=1}^M([n_j]!)^{-\frac
12}(B_j^{\dagger })^{n_j}\mid 0\rangle . 
\]
The notation $[n]!=\prod_{k=1}^n[k]$ is used and the 'box' is 
\[
\lbrack n]=\frac{1-q^{-2n}}{1-q^{-2}},\bl{1} 
[n]!=\prod\limits_{l=1}^{n}[l]. 
\]
The $q-$boson operators (\ref{qbos}) can be expressed in terms of the
ordinary canonical bosons 
\begin{equation}
b_j^{\dagger }\,,\,b_j,[b_i,b_j^{\dagger }]=\delta _{ij},\,N_j=b_j^{\dagger
}b_j,  \label{bos}
\end{equation}
as 
\begin{equation}
B_j=(B_j^{\dagger })^{\dagger }=\sqrt{\frac{[N_j+1]}{N_j+1}}b_j\,.
\label{qbos-bos}
\end{equation}

If $q\rightarrow 1\,(\gamma \rightarrow 0)$, the $q-$bosons become ordinary
bosons 
\[
B_j\rightarrow b_j\,,\,B_j^{\dagger }\rightarrow b_j^{\dagger }. 
\]
At $\gamma =0$ (the free boson limit) the Hamiltonian (\ref{hamgen}) becomes
the linear hopping model Hamiltonian 
\begin{equation}
H_b=-\frac 12\sum_{n=1}^M(b_n^{\dagger }b_{n+1}+b_nb_{n+1}^{\dagger }-2N_n),
\label{hambos}
\end{equation}
while the $q-$boson algebra is the ordinary boson algebra.

In the continuum scaling limit the lattice constant $\delta \rightarrow 0,$ $%
\,M\delta =L,\,\gamma =\frac 12c\delta ,$ and the $q$-boson Hamiltonian (\ref
{hamgen}) is just the Hamiltonian of the Bose gas model with the repulsive $%
\delta $-function interaction of strength $c:$ 
\[
H =\int dx\{\partial _xb^{\dagger }(x)\partial _xb(x)+cb^{\dagger
}(x)b^{\dagger }(x)b(x)b(x)\}, 
\]
\[
\lbrack b(x),b^{\dagger }(y)] =\delta (x-y). 
\]

Representation (\ref{qbos-bos}) shows that (\ref{hamgen}) involves
non-linearities of all orders about the ordinary boson hopping model (\ref
{hambos}). The deformation parameter plays the role of the coupling
constant. Really, for small $\gamma \rightarrow 0$ the hopping model (\ref
{hamgen}) expands to 
\[
H_\gamma =H_b-\frac \gamma 2\sum_{n=1}^M\{b_{n+1}b_n^{\dagger
}N_n+b_{n+1}^{\dagger }(N_n+N_{n+1})b_n+N_{n+1}b_{n+1}b_n^{\dagger }\}+{\cal %
O}(\gamma ^2). 
\]
We can consider the $q$-boson model as a strongly interacting boson model.
On the other hand, the representation (\ref{qbos-bos}) suggests that the
kinetic energy of (\ref{hamgen}) involves boson correlations, i.e., the
hopping terms between adjacent sites depend on the occupation of those
sites. These models are known as the strongly correlated ones.

From this point of view the limit $q\rightarrow \infty \,(\gamma \rightarrow
\infty )$ is of interest. It follows from (\ref{frep}) and (\ref{qbos-bos})
that the operators $B,B^{\dagger }$ transform into operators $\phi ,\phi
^{\dagger }:$%
\begin{equation}
B_j\rightarrow \phi _j=(N_j+1)^{-\frac 12}b_j\,,  \label{phase}
\end{equation}
\[
B_j^{\dagger }\rightarrow \phi _j^{\dagger }=b_j^{\dagger }(N_j+1)^{-\frac
12}, 
\]
with the commutation relations 
\begin{equation}
\lbrack N_i,\phi _j]=-\phi _i\delta _{ij}\,,\,[N_i,\phi _j^{\dagger }]=\phi
_i^{\dagger }\delta _{ij}\,,\,\,[\phi _i,\phi _j^{\dagger }]=\pi _i\delta
_{ij}\,  \label{phalg}
\end{equation}
where $\pi _j$ is the local vacuum projector, $\pi _j=(\mid 0\rangle \langle
0\mid )_j.$ One may verify that $\phi $, $\phi
^{\dagger }$, and $N$ can be expressed in terms of the Fock states, $\mid
n\rangle ,$ as 
\[
\phi =\sum_{n=0}^\infty \mid n\rangle \langle n+1\mid ,\,\phi ^{\dagger
}=\sum_{n=0}^\infty \mid n+1\rangle \langle n\mid ,N=\sum_{n=0}^\infty n\mid
n\rangle \langle n\mid . 
\]
The introduced operator $\phi $ is ''one-sided unitary'' (or isometric),
i.e., 
\[
\phi \phi ^{\dagger }=1, 
\]
but 
\[
\phi ^{\dagger }\phi =1-\mid 0\rangle \langle 0\mid . 
\]
The operators $\phi ,\phi ^{\dagger }$ are studied intensively in quantum
optics in connection with the phase-operator problem (see \cite{carni},\cite
{lynch} and references therein). The introduction of phase variables for
bosons was discussed in \cite{ander}. The relative phase of boson fields is
important in the theory of beam splitters \cite{ls} and Josephson junctions 
\cite{ban}.

For $\gamma =\infty ,$ the Hamiltonian (\ref{hamhim}) becomes 
\begin{equation}
H=-\frac 12\sum_{n=1}^M(\phi _n^{\dagger }\phi _{n+1}+\phi _n\phi
_{n+1}^{\dagger }-2N_n)-\bar \mu \hat N,  \label{ham}
\end{equation}
and 
\[
\lbrack H,\hat N]=0, 
\]
where the total number operator $\hat N$ is given by (\ref{nh}). This model
is called the phase model. It should be mentioned that the case $M=2$ with $%
\bar \mu =0\,$ corresponds to the phase difference operator considered in 
\cite{carni} .

\section{The Bethe Ansatz solution of the model}

\setcounter{equation}{0} Define the $L-$operator for the $q$-bosons hopping
model at lattice site $n$ as

\begin{equation}
L_n(\lambda )=\left( 
\begin{array}{cc}
e^\lambda & \chi B_n^{\dagger } \\ 
\chi B_n & e^{-\lambda }
\end{array}
\right) ,  \label{lo}
\end{equation}
where $B_n,B_n^{\dagger }$ are the $q$-bosons (\ref{qbos}), $\chi =\sqrt{%
1-q^{-2}},$ and $\lambda \in {\cal C}{\bf \ }$is the spectral parameter.
This $L$-operator satisfies the bilinear relation: 
\begin{equation}
R(\lambda ,\mu )L_n(\lambda )\otimes L_n(\mu )=L_n(\mu )\otimes L_n(\lambda
)R(\lambda ,\mu ),  \label{rll}
\end{equation}
with the (gauge transformed) trigonometric $R$-matrix 
\begin{equation}
R(\lambda ,\mu )=\left( 
\begin{array}{cccc}
f(\mu ,\lambda ) & 0 & 0 & 0 \\ 
0 & g(\mu ,\lambda ) & q & 0 \\ 
0 & q^{-1} & g(\mu ,\lambda ) & 0 \\ 
0 & 0 & 0 & f(\mu ,\lambda )
\end{array}
\right) ,  \label{rm}
\end{equation}
with the matrix elements being defined by the functions 
\begin{equation}
f(\lambda ,\mu )=\frac{\sinh (\lambda -\mu +\gamma )}{\sinh (\lambda -\mu )}%
;\,\,g(\lambda ,\mu )=\frac{\sinh \gamma }{\sinh (\lambda -\mu )}.
\label{fg}
\end{equation}
The monodromy matrix is introduced in the usual way as 
\begin{equation}
T(\lambda )=L_M(\lambda )...L_1(\lambda )=\left( 
\begin{array}{cc}
A(\lambda ) & B(\lambda ) \\ 
C(\lambda ) & D(\lambda )
\end{array}
\right) .  \label{mm}
\end{equation}
The 16 commutation relations of its matrix elements are then given by
the $R$-matrix 
\begin{equation}
R(\lambda ,\mu )T(\lambda )\otimes T(\mu )=T(\mu )\otimes T(\lambda
)R(\lambda ,\mu ),  \label{ttr}
\end{equation}
Let us write explicitly the relations important for deriving the Bethe
Ansatz 
\begin{equation}
qA(\lambda )B(\mu )=f(\lambda ,\mu )B(\mu )A(\lambda )+g(\mu ,\lambda
)B(\lambda )A(\mu ),  \label{comr}
\end{equation}
\[
qD(\lambda )B(\mu )=f(\mu ,\lambda )B(\mu )D(\lambda )+g(\lambda ,\mu
)B(\lambda )D(\mu ), 
\]
\[
\lbrack B(\lambda ),B(\mu )]=0, 
\]
\[
C(\lambda )B(\mu )-q^{-2}B(\mu )C(\lambda )=q^{-1}g(\lambda ,\mu
)\{A(\lambda )D(\mu )-A(\mu )D(\lambda )\}. 
\]
The matrix trace of the monodromy matrix is the transfer matrix 
\begin{equation}
\tau (\lambda )=A(\lambda )+D(\lambda ).  \label{trm}
\end{equation}
It follows from (\ref{ttr}) that 
\begin{equation}
\lbrack \tau (\lambda ),\tau (\mu )]=0.  \label{comtt}
\end{equation}
The Hamiltonian (\ref{hamgen}) of the $q$-boson hopping model is expressed
by means of trace identities in terms of the transfer matrix 
\begin{equation}
-2\chi H_q=\frac 12e^{-2\lambda }\frac{\partial e^{M\lambda }\tau (\lambda )%
}{\partial \lambda }\mid _{\lambda =-\infty }-\frac 12e^{2\lambda }\frac{%
\partial e^{-M\lambda }\tau (\lambda )}{\partial \lambda }\mid _{\lambda
=\infty }-2\hat N\chi ^2,  \label{trid}
\end{equation}
and commutes with the transfer matrix 
\begin{equation}
\lbrack H_q,\tau (\lambda )]=0.  \label{ht}
\end{equation}

The eigenvectors of the transfer matrix and hence of the Hamiltonian are of
the form 
\begin{equation}
\mid \psi _N(\lambda _1,...,\lambda _N)\rangle =\prod_{j=1}^NB(\lambda
_j)\mid 0\rangle ,  \label{bvect}
\end{equation}
where the vacuum state $\mid 0\rangle $ is defined in (\ref{qstate}), and
parameters $\lambda _j$ satisfy the Bethe equations: 
\begin{equation}
e^{2M\lambda _j}=\prod_{k\neq j}^N\frac{f(\lambda _k,\lambda _j)}{f(\lambda
_j,\lambda _k)};\,j=1,...,N.  \label{betheeq}
\end{equation}
These states are called Bethe vectors. One can also construct dual Bethe
vectors 
\[
\langle 0|\prod\limits_{j=1}^NC(\lambda _j), 
\]
where $\{\lambda _j\}$ satisfy the same Bethe equation (\ref{betheeq}). The
vacuum vector $|0\rangle $ and the dual vacuum $\langle 0|$ are eigenvectors
of the operators $A(\lambda )$ and $D(\lambda )$, with the ''vacuum''
eigenvalues $a(\lambda )$ and $d(\lambda )$, respectively. For the $q$-boson
hopping model one has 
\begin{equation}
a(\lambda )=e^{M\lambda },\makebox[1em]{}d(\lambda )=e^{-M\lambda }.
\label{ad}
\end{equation}

The eigenvalues $\theta _N$ of the transfer matrix for the eigenvectors (\ref
{bvect}) are 
\[
\tau (\mu )\mid \psi _N\rangle =\theta _N(\mu,\{\lambda_j\})\mid \psi _N\rangle , 
\]
\begin{equation}
q^N\theta _N(\mu,\{\lambda_j\})=e^{M\mu }
\prod_{j=1}^Nf(\lambda _j,\mu )+e^{-M\mu
}\prod_{j=1}^Nf(\mu ,\lambda _j).  \label{eigfun}
\end{equation}
In the explicit form the equations (\ref{betheeq}) are rewritten as 
\begin{equation}
e^{iMp_j}=\prod_{k\neq j}^N\frac{\sin (\frac{p_j-p_k}2+i\gamma )}{\sin (%
\frac{p_j-p_k}2-i\gamma )},  \label{betheeq2}
\end{equation}
where we have introduced the momenta 
\begin{equation}
p=-2i\lambda .  \label{mom}
\end{equation}
Later it will be convenient to use the parameters $\lambda $ for arbitrary
sets $\{\la\}$ and momenta $p$ for the sets satisfying the Bethe equations. In the
scaling limit these equations go into the Bethe equations for the Bose
gas.

Eigenenergies of the Hamiltonian (\ref{hamgen}), $H_q\mid \psi _N\rangle
=E_N\mid \psi _N\rangle ,$ can be found from the equations (\ref{eigfun}), (%
\ref{fg}) and (\ref{trid}) 
\begin{equation}
E_N=\sum_{k=1}^Nh(p_k);\,\,h(p)=2\sin {}^2(p/2).  \label{energy}
\end{equation}
Eigenenergies of the Hamiltonian (\ref{hamhim}), $H_h\mid \psi _N\rangle
=\tilde E_N\mid \psi _N\rangle ,$ are 
\[
\tilde E_N=\sum_{k=1}^N(h(p_k)-\bar \mu ). 
\]

\section{The phase model}

\setcounter{equation}{0} The phase model is the model defined by the
Hamiltonian (\ref{ham}). It belongs to the class of strongly coupled
correlated boson models since the hopping terms between adjacent sites
depend on the occupation of those sites which is evident from the
representation (\ref{phase}) of the $\phi $ -operators in terms of the
ordinary bosons.

The $L-$operator \cite{bik1} of the phase model is obtained by putting $%
\gamma =\infty $ in (\ref{lo}) 
\begin{equation}
L_n(\lambda )=\left( 
\begin{array}{cc}
e^{ip/2} & \phi _n^{\dagger } \\ 
\phi _n & e^{-ip/2}
\end{array}
\right) ,  \label{loph}
\end{equation}
where $\phi _n^{\dagger },\phi _n$ are defined by (\ref{phalg}). This
operator satisfies the bilinear relation 
\[
R(p,s)L_n(p)\otimes L_n(s)=L_n(s)\otimes L_n(p)R(p,s), 
\]
with the 4$\times $4 $R$-matrix $R(p,s)$. The non-zero elements of
the $R$-matrix are 
\[
R_{11}(p,s)=R_{44}(p,s)=f(s,p), 
\]
\[
R_{22}(p,s)=R_{33}(p,s)=g(s,p), 
\]
\[
R_{23}(p,s)=1 
\]
and 
\begin{equation}
f(p,s)=i\frac{e^{i\frac{p-s}2}}{2\sin (\frac{s-p}2)};\,\,g(p,s)=\frac
i{2\sin (\frac{s-p}2)}.  \label{fgph}
\end{equation}
This $R$-matrix is obtained as the limit of the $R$-matrix (\ref{rm}).

The Bethe equations for the model are 
\begin{equation}
\exp \{i(M+N)p_j\}=(-1)^{N-1}\exp \{i\sum_{k=1}^Np_k\},  \label{betheph}
\end{equation}
$(j=1,...,N)$ and are exactly solvable: 
\begin{equation}
p_j=\frac{2\pi I_j+\sum_{k=1}^Np_k}{M+N}\,,  \label{solph}
\end{equation}
where $I_j$ are integers or half-integers depending on $N$ \thinspace being
\thinspace odd or even. The Bethe vectors form a complete orthogonal basis.

The $N$-particle eigenenergies of the Hamiltonian $H-\bar \mu \hat N$ (\ref
{ham}) are 
\begin{equation}
E_N=\sum_{k=1}^N(h(p_k)-\bar \mu );\,\,h(p)=2\sin {}^2(p/2).  \label{enph}
\end{equation}
Here $\bar \mu $ is the chemical potential, $0\leq \bar \mu \leq 1.$ For $%
M=2 $ and $\bar \mu =0$ this result coincides with one obtained in \cite
{carni}.

The thermodynamics of the model is handled in the standard way. It will be
considered for the case of the zero total momentum $P=\sum_{k=1}^Np_k=0.$
The state of the thermal equilibrium of the model
 at finite temperatures $\beta ^{-1}$ is
determined through the solution of the nonlinear integral Yang-Yang equations 
which are drastically symplified in our case
\begin{equation}
\epsilon (p)=h(p)-\bar \mu -(2\pi \beta )^{-1}\int_{-\pi }^\pi \ln
(1+e^{\beta \epsilon (p)})dp,  \label{eenph}
\end{equation}
\[
2\pi \rho (p)(1+e^{\beta \epsilon (p)})=1+\int_{-\pi }^\pi \rho (p)dp. 
\]
The function $\rho (p)$ is a quasi-particle density while $\epsilon (p)$ is
the excitation energy. The pressure is then 
\begin{equation}
{\cal P}=(2\pi \beta )^{-1}\int_{-\pi }^\pi \ln (1+e^{\beta \epsilon (p)})dp,
\label{presph}
\end{equation}
and the density is 
\begin{equation}
D=\frac{\partial {\cal P}}{\partial \bar \mu }=\int_{-\pi }^\pi \rho (p)dp.
\label{denph}
\end{equation}
So we have 
\begin{equation}
\epsilon (p)=h(p)-\bar \mu -{\cal P}  \label{eneph}
\end{equation}
and the quasi-particle density has the Fermi-like distribution 
\begin{equation}
2\pi \rho (p)=(1+D)(1+e^{\beta \epsilon (p)})^{-1}.  \label{disph}
\end{equation}

At zero temperature ($\beta ^{-1}=0$) the ground state is the Fermi zone, $%
-\Lambda \leq p\leq \Lambda $ ($\Lambda \leq \pi ),$ filled by the particles
with the negative energies $\epsilon _0(p).$ The pressure and density are
now 
\begin{equation}
{\cal P}_0=-(2\pi )^{-1}\int_{-\Lambda }^\Lambda \epsilon
_0(p)dp,\,\,D_0=\int_{-\Lambda }^\Lambda \rho (p)dp,  \label{dens0}
\end{equation}
\thinspace where (see (\ref{eneph}) and (\ref{disph})) 
\begin{equation}
\epsilon _0(p)=h(p)-\bar \mu -{\cal P}_0,\,\,\epsilon _0(\pm \Lambda )=0;
\label{Lambda}
\end{equation}
\[
2\pi \rho _0(p)=(1+D_0). 
\]
The bare Fermi momentum $\Lambda $ is expressed as a function of density 
\begin{equation}
\Lambda =\frac{\pi D_0}{1+D_0}.  \label{fermi}
\end{equation}
The Fermi velocity $v$ is given as 
\[
v=\frac{\epsilon _0^{\prime }(\Lambda )}{2\pi \rho _0(\Lambda )}%
=(1+D_0)^{-1}\sin \frac{\pi D_0}{1+D_0}. 
\]

If $\Lambda \rightarrow 0\,(\bar \mu \rightarrow 0),$ then $D_0\rightarrow 0$
and ${\cal P}_0\rightarrow 0$ as should be expected. If $\Lambda \rightarrow
\pi \,(\bar \mu \rightarrow 1)$, all the vacancies are occupied by particles 
$D_0\rightarrow \infty ,\,{\cal P}_0\rightarrow 1$, and the phase model
becomes the classical $XY$ chain in this limit \cite{bbg,bbt}.

\section{Scalar products and norms in the phase model}

\setcounter{equation}{0} Here we calculate the scalar products of the states
produced by the operators $B(\lambda )$ and the norms of Bethe vectors using
the standard procedure \cite{kor1,k}. It is necessary to calculate the
correlation functions. Consider first the scalar products in the $q$-boson
hopping model (\ref{hamgen}) 
\begin{equation}
\tilde S(\{\lambda _j^B\},\{\lambda _k^C\})=\langle 0|C(\lambda _1^C)\dots
C(\lambda _N^C)B(\lambda _N^B)\dots B(\lambda _1^B)|0\rangle ,
\end{equation}
where $C(\lambda ),B(\lambda )$ are the matrix elements of the monodromy
matrix (\ref{mm}) and $\{\lambda _j^B\}$ and $\{\lambda _k^C\}$ are the sets
of arbitrary spectral parameters (Bethe equations are not imposed). Using
the commutation relations (\ref{comr}) one gets 
\begin{equation}
\tilde S(\{\lambda _j^B\},\{\lambda
_k^C\})=\sum\limits_{AD}\prod\limits_{\alpha =1}^Na(\lambda _\alpha
^A)\prod\limits_{\beta =1}^Nd(\lambda _\beta ^D)K_N\left( 
\begin{array}{cc}
\{\lambda ^C\} & \{\lambda ^B\} \\ 
\{\lambda ^A\} & \{\lambda ^D\}
\end{array}
\right) .  \label{sumabcd}
\end{equation}
Here the sum is taken over all the partitions of the set $\{\lambda
_j^B\}\cup \{\lambda _k^C\}$ into two subsets 
\[
\{\lambda _j^B\}\cup \{\lambda _k^C\}=\{\lambda _\alpha ^A\}\cup \{\lambda
_\beta ^D\}. 
\]
The number of elements in each set equals $N$: $\mbox{card}\{\lambda _j^B\}=%
\mbox{card}\{\lambda _k^C\}=\mbox{card}\{\lambda _\alpha ^A\}=\mbox{card}%
\{\lambda _\beta ^D\}=N$. Functions $a(\lambda )$ and $d(\lambda )$ are the
vacuum eigenvalues of the operators $A(\lambda )$ and $D(\lambda )$ (\ref{ad}%
)

The coefficient $K_N\left( 
\begin{array}{cc}
\{\lambda ^C\} & \{\lambda ^B\} \\ 
\{\lambda ^C\} & \{\lambda ^B\}
\end{array}
\right) $ (the ''highest'' coefficient) can be represented in terms of the
partition function of the six-vertex model corresponding to the $R$-matrix (%
\ref{rm}). It was shown in \cite{kit} that this partition function is just
the partition function of the six-vertex model corresponding to the XXZ $R$%
-matrix. The explicit expression for this partition function was given in 
\cite{ick}. This function depends on two sets of parameters $\{\lambda _a\}$
and $\{\nu _k\}$, 
\begin{equation}
Z_N=(-1)^N\frac{\prod\limits_{\alpha =1}^N\prod\limits_{k=1}^N\sinh (\lambda
_a-\nu _k-\frac \gamma 2)\sinh (\lambda _a-\nu _k+\frac \gamma 2)}{%
\prod\limits_{\alpha <\beta }\sinh (\lambda _a-\lambda
_b)\prod\limits_{k<l}\sinh (\nu _l-\nu _k)}\mbox{det}_N{\cal M},
\label{partfun2}
\end{equation}
where the $N\times N$ matrix ${\cal M}$ has the following form 
\begin{equation}
{\cal M}_{\alpha ,k}=\frac{\sinh \gamma }{\sinh (\lambda _a-\nu _k-\frac
\gamma 2)\sinh (\lambda _a-\nu _k+\frac \gamma 2)}.  \label{partfun3}
\end{equation}
The highest coefficient can be expressed as follows 
\begin{equation}
K_N\left( 
\begin{array}{cc}
\{\lambda ^C\} & \{\lambda ^B\} \\ 
\{\lambda ^C\} & \{\lambda ^B\}
\end{array}
\right) =e^{-2\gamma N^2}\left( \prod\limits_{j,k=1}^N\sinh (\lambda
_j^B-\lambda _k^C)\right) ^{-1}Z_N(\{\lambda _j^B\},\{\lambda _k^C+\frac
\gamma 2\}).  \label{qhighK}
\end{equation}
To prove this proposition one should consider the inhomogeneous gauge
transformed XXZ model on a lattice of $N$ sites taking into account that the
highest coefficient depends only on the $R$-matrix (see \cite{k}). Other
coefficients in (\ref{sumabcd}) can be analogously represented as 
\[
K_N\left( 
\begin{array}{cc}
\{\lambda ^C\} & \{\lambda ^B\} \\ 
\{\lambda ^A\} & \{\lambda ^D\}
\end{array}
\right) =e^{-2\gamma n_0(N-n_0)}\left( \prod\limits_{j\in
AC}\prod\limits_{k\in DC}f(\lambda _j^{AC},\lambda _k^{DC})\right) \times 
\]
\begin{equation}
\left( \prod\limits_{l\in AB}\prod\limits_{m\in DB}f(\lambda _l^{AB},\lambda
_m^{DB})\right) K_{n_0}\left( 
\begin{array}{cc}
\{\lambda ^{AB}\} & \{\lambda ^{DC}\} \\ 
\{\lambda ^{AB}\} & \{\lambda ^{DC}\}
\end{array}
\right) K_{N-n_0}\left( 
\begin{array}{cc}
\{\lambda ^{AC}\} & \{\lambda ^{DB}\} \\ 
\{\lambda ^{AC}\} & \{\lambda ^{DB}\}
\end{array}
\right) .  \label{qallK}
\end{equation}
Now it is not difficult to represent the scalar products for the q-boson
hopping model as mean values of determinants depending on dual fields using
the approach of \cite{kor}.

In this paper we will not, however, use this method concentrating our
attention on the phase model only. It appears that the scalar products and
correlation functions in the phase model can be represented without
auxiliary fields. The matrix elements of the matrix 
${\cal M}$ (\ref{partfun3})
the limit $\gamma
\rightarrow \infty $ are given as 
\[
{\cal M}_{jk}=\frac{2e^{2\lambda _j^B}}{e^{2\lambda _k^C}-e^{2\lambda _j^B}}%
+O(e^{-\gamma }), 
\]
\[
\mbox{det}{\cal M}=2^N\frac{\prod\limits_{j<l}(e^{2\lambda _j^B}-e^{2\lambda
_l^B})\prod\limits_{k<m}(e^{2\lambda _m^C}-e^{2\lambda _k^C})}{%
\prod\limits_{j,k}(e^{2\lambda _j^B}-e^{2\lambda _k^C})}e^{2\sum_{j=1}^N%
\lambda _j^B}+O(e^{-\gamma }), 
\]
for the highest coefficient (\ref{qhighK}) we obtain 
\[
K_N\left( 
\begin{array}{cc}
\{\lambda ^C\} & \{\lambda ^B\} \\ 
\{\lambda ^C\} & \{\lambda ^B\}
\end{array}
\right) =\left( \prod\limits_{j<l}(e^{2\lambda _j^B}-e^{2\lambda
_l^B})\prod\limits_{k<m}(e^{2\lambda _m^C}-e^{2\lambda _k^C})\right)
^{-1}\times 
\]
\begin{equation}
\times \mbox{det}_N{\cal K}(\{\lambda _j^B\},\{\lambda _k^C\}),
\label{highKdet}
\end{equation}
\[
{\cal K}_{jk}=\frac{\exp ((2N-1)\lambda _k^C+\lambda _j^B)}{e^{2\lambda
_k^C}-e^{2\lambda _j^B}}, 
\]
and other coefficients (\ref{qallK}) are 
\[
K_N\left( 
\begin{array}{cc}
\{\lambda ^C\} & \{\lambda ^B\} \\ 
\{\lambda ^A\} & \{\lambda ^D\}
\end{array}
\right) =\left( \prod\limits_{j<l}(e^{2\lambda _j^B}-e^{2\lambda
_l^B})\prod\limits_{k<m}(e^{2\lambda _m^C}-e^{2\lambda _k^C})\right)
^{-1}\times 
\]
\begin{equation}
\times (-1)^{([P]+[Q])}\mbox{det}_{n_0}{\cal K}(\{\lambda ^{DC}\},\{\lambda
^{AB}\})\mbox{det}_{N-n_0}{\cal K}(\{\lambda ^{DB}\},\{\lambda ^{AC}\}).
\label{allKdet}
\end{equation}
Here $P$ and $Q$ are the transpositions $\{\lambda ^B\}\rightarrow \{\lambda
^{AB}\}\cup \{\lambda ^{DB}\},\{\lambda ^C\}\rightarrow \{\lambda
^{AC}\}\cup \{\lambda ^{DC}\}$ respectively.

It is convenient to consider the ''normalized'' scalar product in the phase
model 
\begin{equation}
S(\{\lambda _j^B\},\{\lambda _k^C\})=\langle 0|\prod\limits_{k=1}^N{\bf C}%
(\lambda _k^C)\prod\limits_{j=1}^N{\bf B}(\lambda _j^B)|0\rangle ,
\end{equation}
where 
\begin{equation}
{\bf B}(\lambda _j^B)=\frac{B(\lambda _j^B)}{d(\lambda _j^B)},\makebox[1em]{}%
{\bf C}(\lambda _k^C)=\frac{C(\lambda _k^C)}{d(\lambda _k^C)}.  \label{bold}
\end{equation}
From (\ref{sumabcd},\ref{allKdet}) one obtains 
\[
S(\{\lambda _j^B\},\{\lambda _k^C\})=\left( \prod\limits_{j<l}(e^{2\lambda
_j^B}-e^{2\lambda _l^B})\prod\limits_{k<m}(e^{2\lambda _m^C}-e^{2\lambda
_k^C})\right) ^{-1}\times 
\]
\[
\times \sum\limits_{A,D}(-1)^{([P]+[Q])}\left( \prod\limits_{l=1}^Nr(\lambda
_l^A)\right) \mbox{det}_{n_0}{\cal K}(\{\lambda ^{DC}\},\{\lambda ^{AB}\})%
\mbox{det}_{N-n_0}{\cal K}(\{\lambda ^{DB}\},\{\lambda ^{AC}\}), 
\]
where $r(\lambda )=a(\lambda )/d(\lambda )$. Using the formula for the
determinant of the sum of two matrices we can express the scalar product 
as a determinant of an $N\times N$ matrix 
\begin{equation}
S(\{\lambda _j^B\},\{\lambda _k^C\})=\left( \prod\limits_{j<l}(e^{2\lambda
_j^B}-e^{2\lambda _l^B})\prod\limits_{k<m}(e^{2\lambda _m^C}-e^{2\lambda
_k^C})\right) ^{-1}\mbox{det}_NF(\{\lambda _j^B\},\{\lambda _k^C\}),
\label{scprod}
\end{equation}
\begin{equation}
F_{jk}=\frac{r(\lambda _j^B)\exp ((2N-1)\lambda _j^B+\lambda _k^C)-r(\lambda
_k^C)\exp ((2N-1)\lambda _k^C+\lambda _j^B)}{e^{2\lambda _j^B}-e^{2\lambda
_k^C}}.
\end{equation}
For the phase model, $r(\lambda )=e^{2\lambda M}$ (\ref{ad}). Hence we
obtain 
\begin{equation}
F_{jk}=\frac{\exp ((2N+2M-1)\lambda _j^B+\lambda _k^C)-\exp
((2N+2M-1)\lambda _k^C+\lambda _j^B)}{e^{2\lambda _j^B}-e^{2\lambda _k^C}}.
\label{matrsp}
\end{equation}
There is an important case when $\{\lambda ^B\}=\{\lambda ^C\}$. The
diagonal elements of the matrix $F$ are then 
\[
F_{jj}=(M+N-1)e^{2(M+N-1)\lambda _j}, 
\]

Turn now to the norms of Bethe eigenvectors 
\begin{equation}
|\psi (p_1,\dots ,p_N)\rangle _N=\prod\limits_{j=1}^N{\bf B}(\lambda
_j)|0\rangle ,\makebox[1em]{}\langle \psi (p_1,\dots ,p_N)|=\langle
0|\prod\limits_{j=1}^N{\bf C}(\lambda _j),\makebox[1em]{}p_j=-2i\lambda _j,
\label{bstates}
\end{equation}
where $\{p\}$ is a solution of the Bethe equations (\ref{betheph}).Now we
have 
\[
F_{jj}=(M+N-1)e^{i(P-p_j)},\makebox[2em]{}P=\sum\limits_{j=1}^Np_j, 
\]
\[
F_{jk}=-e^{i(P-\frac{p_j}2-\frac{p_k}2)},\bl{1}j\neq k. 
\]
It is not difficult to calculate the determinant 
\[
\mbox{det}F=e^{iP(N-1)}M(M+N)^{N-1}, 
\]
and for the norm of any Bethe eigenstate we obtain the formula of the Gaudin
type 
\begin{equation}
{\cal N}^2(p_1,\dots ,p_N)=\langle \psi (p_1,\dots ,p_N)|\psi (p_1,\dots
,p_N)\rangle =\left( \prod\limits_{j\neq k}\frac 1{e^{ip_j}-e^{ip_k}}\right)
e^{iP(N-1)}M(M+N)^{N-1}.  \label{gaudin}
\end{equation}

\section{The darkness formation probability}

\setcounter{equation}{0} Here we will calculate the darkness formation
probability, i.e., the probability of the states with no
particles on the first $m$ sites of the lattice. Formally it can be defined
as the normalized mean value
\begin{equation}
\tau (m,\{p\})=\left. {\cal N}^{-2}(p_1,\dots ,p_N)\langle \psi (p_1,\dots
,p_N)|\exp \{\alpha Q(m)\}|\psi (p_1,\dots ,p_N)\rangle \right| _{\alpha
=-\infty },  \label{dfkdef}
\end{equation}
where $Q(m)=\sum_{j=1}^mN_j$ is the number of particles operator on the
first $m$ sites.

The monodromy matrix can be represented in the following form 
\[
T(\lambda )=T_2(\lambda )T_1(\lambda ), 
\]
\begin{equation}
T_2(\lambda )=L_M(\lambda )\dots L_{m+1}(\lambda )=\left( 
\begin{array}{cc}
A_2(\lambda ) & B_2(\lambda ) \\ 
C_2(\lambda ) & D_2(\lambda )
\end{array}
\right) ,  \label{split}
\end{equation}
\[
T_1(\lambda )=L_m(\lambda )\dots L_1(\lambda )=\left( 
\begin{array}{cc}
A_1(\lambda ) & B_1(\lambda ) \\ 
C_1(\lambda ) & D_1(\lambda )
\end{array}
\right). 
\]
The bare vacuum (\ref{qstate}) can be represented as 
\[
|0\rangle =|0\rangle _{{\rm II}}\otimes |0\rangle _{{\rm I}}, 
\]
where $|0\rangle _{{\rm I}}$ and $|0\rangle _{{\rm II}}$ are the Fock vacua
for the monodromy matrices $T_1(\lambda )$ and $T_2(\lambda )$,
respectively. We denote $r_1(\lambda )$ and $r_2(\lambda )$ 
the ratios of the corresponding vacuum eigenvalues
\[r_1(\lambda )=a_1(\lambda )/d_1(\lambda )=\exp
\{2m\lambda \}\bl{1},r_2(\lambda )=a_2(\lambda )/d_2(\lambda )=\exp
\{2(M-m)\lambda \}.
\]
The numerator of (\ref{dfkdef}) is a particular value of the following
matrix element 
\begin{equation}
T(\{\lambda _j^B\},\{\lambda _k^C\})=\langle 0|\prod\limits_{k=1}^N{\bf C}%
(\lambda _k^C)\exp \{\alpha Q(m)\}\prod\limits_{j=1}^N{\bf B}(\lambda
_j^B)|0\rangle |_{\alpha =-\infty },  \label{efp}
\end{equation}
where the operators ${\bf B}(\lambda )$ and ${\bf C}(\lambda )$ are defined
by (\ref{bold}) and $\{\lambda _j^B\},\{\lambda _k^C\}$ are arbitrary sets
of $N$ spectral parameters. Using the decomposition (\ref{split}) one
obtains 
\[
T(\{\lambda _j^B\},\{\lambda _k^C\})=\prod\limits_{j=1}^Nr_1(\lambda _j^C)\vphantom{\langle}_{%
{\rm II}}\langle 0|\prod\limits_{k=1}^N{\bf C_2}(\lambda
_k^C)\prod\limits_{j=1}^N{\bf B_2}(\lambda _j^B)|0\rangle _{{\rm II}}. 
\]
The scalar product here can be calculated by means of (\ref{scprod}), with
the result 
\begin{equation}
T(\{\lambda _j^B\},\{\lambda _k^C\})=\left( \prod\limits_{j<l}(e^{2\lambda
_j^B}-e^{2\lambda _l^B})\prod\limits_{k<m}(e^{2\lambda _m^C}-e^{2\lambda
_k^C})\right) ^{-1}\mbox{det}{\cal T}(\{\lambda _j^B\},\{\lambda _k^C\}),
\label{efp1}
\end{equation}
the matrix elements of the $N\times N$ matrix ${\cal T}$ being equal to 
\[
{\cal T}_{jk}=\frac 1{e^{2\lambda _k^C}-e^{2\lambda _j^B}}\times 
\]
\[
\times \left( \exp ((2N+2M-1)\lambda _k^C+\lambda _j^B)-\exp
((2N+2M-2m-1)\lambda _j^B+(2m+1)\lambda _k^C)\right) . 
\]
If the sets coincide, $\{\lambda ^B\}=\{\lambda ^C\}=\{\lambda \}$, then the
diagonal elements should be understood in the sense of the l'H\^opital rule 
\[
{\cal T}_{jj}=(M+N-m-1)e^{2(M+N-1)\lambda _j}. 
\]
If the spectral parameters satisfy the Bethe equations the matrix elements $%
{\cal T}_{jk}$ can be represented in the form ($p_j=-2i\lambda _j$ and $P$
is the sum of momenta) 
\[
{\cal T}_{jj}=(M+N-m-1)e^{i(P-p_j)}, 
\]
\[
{\cal T}_{jk}=e^{i(P-p_k)}e^{i\frac{m+1}2(p_j-p_k)}\frac{\sin \frac{m+1}%
2(p_j-p_k)}{\sin \frac 12(p_j-p_k)},\makebox[1em]{}j\neq k. 
\]
Thus one obtains the following result for the darkness formation probability 
\begin{eqnarray}
\tau (m,\{p\}) &=&\frac{T(\{p\},\{p\})}{{\cal N}^2(\{p\})}=(1+D)\mbox{det}%
\tilde T(m,\{p\}),  \label{efp2} \\
\tilde T_{jk} &=&\delta _{j,k}-\frac 1M\frac 1{1+D}\frac{\sin \frac{m+1}%
2(p_j-p_k)}{\sin \frac 12(p_j-p_k)},
\end{eqnarray}
where $D=N/M$.

\section{Time-dependent correlation functions}

\setcounter{equation}{0} In this section a determinant representation for
the two-point time-dependent correlation function of fields on the finite
lattice is obtained. The derivation is somewhat similar to the derivation in
the case of the XX0 chain \cite{cikt2}. To this end one needs first a
representation for the form factors 
\begin{equation}
G_N(m,\{p\},\{q\})=\frac{\langle \psi (q_1,\dots ,q_{N-1})|\phi _m|\psi
(p_1,\dots ,p_N)\rangle }{{\cal N}(q_1,\dots ,q_{N-1}){\cal N}(p_1,\dots
,p_N)},  \label{corff1}
\end{equation}
\begin{equation}
G_N^{(+)}(m,\{p\},\{q\})=\frac{\langle \psi (q_1,\dots ,q_{N+1})|\phi
_m^{\dagger }|\psi (p_1,\dots ,p_N)\rangle }{{\cal N}(q_1,\dots ,q_{N+1})%
{\cal N}(p_1,\dots ,p_N)},  \label{corff2}
\end{equation}
where $|\psi (\{p\})\rangle $ are Bethe vectors and ${\cal N}(\{p\})$ are
their norms  (\ref{gaudin})
\[
{\cal N}(\{p\})=\langle \psi (\{p\})|\psi (\{p\})\rangle ^{1/2}, 
\]

The determinant representations for the form factors of the phase model were
obtained in \cite{kit} (a brief derivation of these results is given in
Appendix A): 
\[
G_N(m,\{p\},\{q\})=M^{-1}(M+N)^{-N+\frac 32}\tilde Ze^{-im\Theta }\left( 
\frac{M+N}{M+N-1}\right) ^{\frac N2-1}\times 
\]
\begin{equation}
\times (1+\frac \partial {\partial x})\mbox{det}_{N-1}(D_1-xD_2)|_{x=0},
\label{difff}
\end{equation}
where  is a complex number depending on $P$ and $Q$, $|\tilde Z|=1$%
The explicit form of $\tilde Z$ is not written here since since it 
appears to be of no importance for calculating the correltors. 
We use convenient notations: 
\[
\Theta =Q-P,\makebox[1em]{}Q=\sum\limits_jq_k,\makebox[1em]{}%
P=\sum\limits_jp_j. 
\]
Matrix elements of matrices $D_1$ and $D_2$ are 
\begin{eqnarray}
D_{1_{jk}} &=&e^{\frac i2p_j}(\frac{\cos \frac 12(\Theta -p_j)}{\tan \frac
12(p_j-q_k)}+\sin \frac 12(\Theta -p_j)),  \label{d1d2} \\
D_{2_{jk}} &=&e^{\frac i2p_N}(\frac{\cos \frac 12(\Theta -p_N)}{\tan \frac
12(p_N-q_k)}+\sin \frac 12(\Theta -p_N)).
\end{eqnarray}
Analogously, for the form factor (\ref{corff2}) we have (the star denotes the
complex conjugation)
\[
G_N^{(+)}(m,\{p\},\{q\})=M^{-1}(M+N)^{-N+\frac 12}\left( \frac{M+N+1}{M+N}%
\right) ^{\frac N2}\tilde Z^{*}e^{im\Theta }\times 
\]
\begin{equation}
\times (1+\frac \partial {\partial x})\mbox{det}%
_N(D_1^{(+)}-xD_2^{(+)})|_{x=0},  \label{ff+}
\end{equation}
where 
\begin{eqnarray}
D_{1_{jk}}^{(+)} &=&e^{-\frac i2q_k}(\frac{\cos \frac 12(\Theta -q_k)}{\tan
\frac 12(q_k-p_j)}+\sin \frac 12(\Theta -q_k)),  \label{d1+d2+} \\
D_{2_{jk}}^{(+)} &=&e^{-\frac i2q_{N+1}}(\frac{\cos \frac 12(\Theta -q_{N+1})%
}{\tan \frac 12(q_{N+1}-p_j)}+\sin \frac 12(\Theta -q_{N+1})).
\end{eqnarray}

Consider now the normalized mean value of the time-dependent product of two
phase operators 
\begin{equation}
f_N^{+}(\{p\},m,t)=\frac{\langle 0|\prod\limits_{k=1}^N{\bf C}(p_k)\phi
_{m+1}(t)\phi _1^{\dagger }(0)\prod\limits_{k=1}^N{\bf B}(p_k)|0\rangle }{%
\langle 0|\prod\limits_{k=1}^N{\bf C}(p_k)\prod\limits_{k=1}^N{\bf B}%
(p_k)|0\rangle },  \label{corr}
\end{equation}
where 
\[
\phi _m(t)=\exp [iHt]\phi _m\exp [-iHt], 
\]
with the Hamiltonian $H$ (\ref{ham}). Using the formulae (\ref{difff}), (\ref
{ff+}) we can represent this correlator as follows 
\[
f_N^{+}(\{p\},m,t)=M^{-2}(M+N+1)^{-2N+1}\left( \frac{M+N+1}{M+N}\right)
^{N-1}\times 
\]
\begin{equation}
\times \sum\limits_{\{q\}}\exp \left(
im(\sum\limits_{k=1}^{N+1}q_k-\sum\limits_{j=1}^Np_j)+it(\sum\limits_{j=1}^N%
\varepsilon (p_j)-\sum\limits_{k=1}^{N+1}\varepsilon (q_k))\right) \times
\label{corsum}
\end{equation}
\[
\times \left| (1+\frac \partial {\partial x})\mbox{det}_N(D_1^{(+)}(\{p\},%
\{q\})-xD_2^{(+)}(\{p\},\{q\}))|_{x=0}\right| ^2, 
\]
where $\varepsilon (p)$ is the energy of quasiparticle $\varepsilon
(p)=2\sin ^2\frac p2-\bar {\mu }$ (\ref{enph}).

The summation in (\ref{corsum}) is taken over all the solutions $\{q\}$ of
the Bethe equations (\ref{betheph}) such that $\mbox{card}\{q\}=N+1$. It can
be seen from (\ref{solph}) that $Q\equiv \sum q_j=\frac{2\pi K}M$ where $K$
is an integer, $-\frac M2<K\leq \frac M2$ . Then the sum over all the
solutions of the Bethe equation can be rewritten as the sum over all such $Q$
and the sum over all the sets of $N+1$ different $q_k$ satisfying the following
conditions 
\[
q_k=\frac{2\pi I_k+Q}{M+N+1},\makebox[1em]{}-\pi <q_k\leq \pi ,%
\makebox[1em]{}Q=\sum\limits_{k=1}^{N+1}q_K. 
\]
Taking into account that the form factor is an antisymmetric function of
momenta $\{q\}$ one can make the following substitution 
\[
\sum\limits_{\{q\}}\dots \longrightarrow \frac
1{(M+N+1)(N+1)!}\sum\limits_Q\sum\limits_{l=0}^{M+N}\sum\limits_{q_1}\dots
\sum\limits_{q_{N+1}}\exp \left( -il(Q-\sum\limits_{k=1}^{N+1}q_k)\right)
\dots 
\]
The determinants in (\ref{corsum}) can be then rewritten as follows 
\[
\left| (1+\frac \partial {\partial x})\mbox{det}_N(D_1^{(+)}(\{p\},\{q%
\})-xD_2^{(+)}(\{p\},\{q\}))|_{x=0}\right| ^2= 
\]
\[
=\sum\limits_{{\cal Q}}(-1)^{[{\cal Q}]}\prod\limits_{a=1}^Ne^{\frac i2q_{%
{\cal Q}_a}}(\frac{\cos \frac 12(\Theta -q_{{\cal Q}_a})}{\tan \frac 12(q_{%
{\cal Q}_a}-p_a)}+\sin \frac 12(\Theta -q_{{\cal Q}_a}))\times 
\]
\[
\times (1+\frac \partial {\partial x})\mbox{det}_N(D_1^{(+)}(\{p\},\{q%
\})-xD_2^{(+)}(\{p\},\{q\}))|_{x=0}, 
\]
where the sum is taken over all the permutations ${\cal Q}$ of $\{q_1,\dots
,q_{N+1}\}.$ We can perform this summation using again the antisymmetry of
the form factor 
\[
\frac 1{(N+1)!}\sum\limits_{{\cal Q}}(-1)^{[{\cal Q}]}\prod%
\limits_{a=1}^Ne^{\frac i2q_{{\cal Q}_a}}(\frac{\cos \frac 12(\Theta -q_{%
{\cal Q}_a})}{\tan \frac 12(q_{{\cal Q}_a}-p_a)}+\sin \frac 12(\Theta -q_{%
{\cal Q}_a}))\longrightarrow 
\]
\[
\longrightarrow \prod\limits_{a=1}^Ne^{\frac i2q_a}(\frac{\cos \frac
12(\Theta -q_a)}{\tan \frac 12(q_a-p_a)}+\sin \frac 12(\Theta -q_a)). 
\]

Taking into account that $\mbox{det}(D_1-xD_2)$ is a linear function of $x$
it is possible to rewrite (\ref{corsum}) as 
\begin{equation}
f_N^{+}(\{p\},m,t)=\left( \frac{M+N+1}{M+N}\right)
^{N-1}\sum\limits_{l=0}^{M+N}\frac{M+N+1}{M^2}\sum\limits_\Theta
e^{i(m-l)\Theta }h_N^{+}(\{p\},l,t,\Theta ),  \label{cor1}
\end{equation}
\[
h_N^{+}(\{p\},l,t,\Theta )=\exp \left(
-il\sum\limits_{j=1}^Np_j+it\sum\limits_{j=1}^N\varepsilon (p_j)\right)
\times 
\]
\begin{equation}
\times \sum\limits_{q_1}\dots \sum\limits_{q_{N+1}}\left( \frac 1{M+N+1}\exp
[ilq_{N+1}-it\varepsilon (q_{N+1})]+\frac \partial {\partial x}\right) %
\mbox{det}W(x)|_{x=0},  \label{hcor}
\end{equation}
where the $N\times N$ matrix $W(x)$ is defined as 
\begin{equation}
W(z)=W^{(1)}-\frac x{M+N+1}W^{(2)},  \label{Wcor1}
\end{equation}
\[
W_{ab}^{(1)}=\frac 1{(M+N+1)^2}\exp [ilq_a-it\varepsilon (q_a)]\left( \frac{%
\cos \frac 12(\Theta -q_a)}{\tan \frac 12(q_a-p_a)}+\sin \frac 12(\Theta
-q_a)\right) \times 
\]
\begin{equation}
\times \left( \frac{\cos \frac 12(\Theta -q_a)}{\tan \frac 12(q_a-p_b)}+\sin
\frac 12(\Theta -q_a)\right) ,  \label{Wcor2}
\end{equation}
\[
W_{ab}^{(2)}=\frac 1{(M+N+1)^2}\exp [i(l+\frac 12)q_a-it\varepsilon
(q_a)]\exp [i(l-\frac 12)q_{N+1}-it\varepsilon (q_{N+1})]\times 
\]
\begin{equation}
\times \left( \frac{\cos \frac 12(\Theta -q_a)}{\tan \frac 12(q_a-p_a)}+\sin
\frac 12(\Theta -q_a)\right) \left( \frac{\cos \frac 12(\Theta -q_{N+1})}{%
\tan \frac 12(q_{N+1}-p_b)}+\sin \frac 12(\Theta -q_{N+1})\right) .
\label{Wcor3}
\end{equation}
The matrix $W^{(2)}$ is of rank one and hence $\mbox{det}_NW(x)$ is a linear
function of $x$.

Let us introduce the following functions 
\begin{equation}
g(l,t)=\frac 1{M+N+1}\sum\limits_q\exp [ilq-it\varepsilon (q)],  \label{g}
\end{equation}
\begin{equation}
e(l,t,p)=\frac 1{M+N+1}\sum\limits_q\frac{\exp [ilq-it\varepsilon (q)]}{\tan
\frac 12(q-p)},  \label{e}
\end{equation}
\begin{equation}
d(l,t,p)=\frac 1{(M+N+1)^2}\sum\limits_q\frac{\exp [ilq-it\varepsilon (q)]}{%
\sin ^2\frac 12(q-p)}.  \label{d}
\end{equation}
The summation in these formulae is made over all the permitted values of $q$
for $\Theta $ fixed, thus these three functions depend also on $\Theta $.
For $t=0$ they can be calculated explicitly. 
\begin{equation}
g(l,0)=\delta _{l,0},  \label{g0}
\end{equation}
\begin{equation}
e(l,0,p)=i(1-\delta _{l,0}-i\tan \frac 12(p-\Theta ))e^{ilp},  \label{e0}
\end{equation}
\begin{equation}
d(l,0,p)=\cos ^{-2}\frac 12(\Theta -p)e^{ilp}+\frac 2{M+N+1}\frac \partial
{\partial p}e(l,0,p).  \label{d0}
\end{equation}
Using these functions we can perform the summations in (\ref{hcor}), bearing
in mind also the following relation 
\[
\cot \frac 12(q-p_a)\cot \frac 12(q-p_b)=\cot \frac 12(p_a-p_b)\left[ \cot
\frac 12(q-p_a)-\cot \frac 12(q-p_b)\right] -1. 
\]
Performing the summation in each row and taking into account that $%
\mbox{det}W(x)$ is a linear function of $x$ we have 
\begin{equation}
h_N^{+}(\{p\},l,t,\Theta )=\left( g(m,t)+\frac \partial {\partial x}\right) %
\mbox{det}_N\left[ S-xR^{+}\right] |_{x=0}.  \label{hcor1}
\end{equation}
The matrices $S$ and $R^{+}$ are given as 
\[
S_{ab}=\frac 1{M+N+1}\left\{ \frac 1{\tan \frac 12(p_a-p_b)}\left( \frac
12(e_{+}^{+}(l,t,p_a,\Theta )+e_{+}^{-}(l,t,p_a,\Theta
))e_{-}(l,t,p_b)-\right. \right. 
\]
\[
\left. -\frac 12(e_{+}^{+}(l,t,p_b,\Theta )+e_{+}^{-}(l,t,p_b,\Theta
))e_{-}(l,t,p_a)\right) - 
\]
\[
-g(l,t)e_{-}(l,t,p_a)e_{-}(l,t,p_b)+\frac i2(e_{+}^{+}(l,t,p_a,\Theta
)-e_{+}^{-}(l,t,p_a,\Theta ))e_{-}(l,t,p_b)+ 
\]
\begin{equation}
\left. \vphantom{\frac{1}{\tan \frac 12(p_a-p_b)}\left(\frac
12(e_{+}^{+}(l,t,p_a,\Theta )\right.}+\frac i2(e_{+}^{+}(l,t,p_b,\Theta
)-e_{+}^{-}(l,t,p_b,\Theta ))e_{-}(l,t,p_a)\right\} ,\makebox[2em]{}{\rm for}%
\makebox[1em]{}a\neq b,  \label{Scor}
\end{equation}
\[
S_{aa}=\frac 12\left( d(l,t,p_a,\Theta )+\frac 12(e^{-i\Theta
}d(l+1,t,p_a,\Theta )+e^{i\Theta }d(l-1,t,p_a,\Theta ))\right) \times 
\]
\[
\times e_{-}(l,t,p_a)e_{-}(l,t,p_a)+\frac
1{M+N+1}\{-g(l,t)e_{-}(l,t,p_a)e_{-}(l,t,p_a)+ 
\]
\begin{equation}
+i(e_{+}^{+}(l,t,p_a,\Theta )-e_{+}^{-}(l,t,p_a,\Theta ))e_{-}(l,t,p_a)\},
\label{Sdiag}
\end{equation}
\begin{equation}
R_{ab}^{+}=\frac 1{M+N+1}e_{+}^{+}(l,t,p_a,\Theta )e_{+}^{-}(l,t,p_b,\Theta
).  \label{Rcor}
\end{equation}
The functions $e_{-}(l,t,p),e_{+}^{+}(l,t,p,\Theta ),e_{+}^{-}(l,t,p,\Theta
) $ are defined as 
\begin{equation}
e_{-}(l,t,p)=\exp \left( -i\frac l2p+i\frac t2\varepsilon (p)\right) ,
\label{e-}
\end{equation}
\[
e_{+}^{+}(l,t,p,\Theta )=\frac 12e_{-}(l,t,p)\times 
\]
\begin{equation}
\times \left( (e(l,t,p)+e^{-i\Theta }e(l+1,t,p))-i(g(l,t)-e^{-i\Theta
}g(l+1,t))\right) ,  \label{e++}
\end{equation}
\[
e_{+}^{-}(l,t,p,\Theta )=\frac 12e_{-}(l,t,p)\times 
\]
\begin{equation}
\times \left( (e(l,t,p)+e^{i\Theta }e(l-1,t,p))+i(g(l,t)-e^{i\Theta
}g(l-1,t))\right) .  \label{e+-}
\end{equation}
In the equal-time case one has explicit expressions for these functions 
\begin{equation}
e_{-}(l,0,p)=e^{-i\frac l2p},  \label{e-0}
\end{equation}
\begin{equation}
e_{+}^{+}(l,0,p,\Theta )=i(1-\delta _{l,0})e^{i\frac l2p},  \label{e++0}
\end{equation}
\begin{equation}
e_{+}^{-}(l,0,p,\Theta )=i(1-\delta _{l,1})e^{i(\Theta -p)}e^{i\frac l2p}.
\label{e+-0}
\end{equation}

Equations (\ref{cor1}) and (\ref{hcor1}-\ref{Rcor}) give a determinant
representation for the correlation function (\ref{corr}) on the finite
lattice.

The calculation of the two-point time-dependent correlation function 
\begin{equation}
f_N^{-}(\{p\},m,t)=\frac{\langle 0|\prod\limits_{k=1}^N{\bf C}(p_k)\phi
_{m+1}^{\dagger }(t)\phi _1(0)\prod\limits_{k=1}^N{\bf B}(p_k)|0\rangle }{%
\langle 0|\prod\limits_{k=1}^N{\bf C}(p_k)\prod\limits_{k=1}^N{\bf B}%
(p_k)|0\rangle },  \label{corr-}
\end{equation}
is quite similar and we give only the final result: 
\begin{equation}
f_N^{-}(\{p\},m,t)=\left( \frac{M+N-1}{M+N}\right)
^{N-1}\sum\limits_{l=1}^{M+N-1}\frac{M+N-1}{M^2}\sum\limits_\Theta
e^{i(m-l)\Theta }h_N^{-}(\{p\},l,t,\Theta ),  \label{cor1-}
\end{equation}
\begin{equation}
h_N^{-}(\{p\},l,t,\Theta )=\frac \partial {\partial x}\mbox{det}_N\left[
S+xR^{-}\right] |_{x=0},  \label{hcor1-}
\end{equation}
\begin{equation}
R_{ab}^{-}=\frac 1{M+N-1}e_{-}(l,t,p_a)e_{-}(l,t,p_b).  \label{Rcor-}
\end{equation}
One should note that in this case the factor $M+N+1$ entering the definitions 
of the matrix $S$ (\ref{Scor},\ref{Sdiag})
and the functions $e$, $d$ and $g$ (\ref{g}-\ref{d}) 
should be replaced by $M+N-1$.

\section{The thermodynamic limit}

\setcounter{equation}{0} Let us consider the thermodynamic limit
(the total number of
sites $M\rightarrow \infty ,$ number of particles $N\rightarrow \infty ,$
the density $D=N/M$ remains finite)  of the correlation functions obtained
in the previous sections. In this limit one should replace all
the sums over momenta by the integrals and the determinants of $N\times N$
matrices by the Fredholm determinants of the corresponding integral operators
acting on the functions on the interval $[-\pi ,\pi ]$ \cite{k}.

The result for the darkness formation probability (\ref{efp2}) is 
\begin{equation}
\tau (m)=(1+D)\mbox{det}(\hat I-\hat T),  \label{efptherm}
\end{equation}
where $\hat I$ is the identity operator and $\hat T$ is an integral operator 
\[
(\hat Tf)(p)=\int\limits_{-\pi }^\pi T(p,q)f(q)dq,
\]
with the kernel 
\begin{equation}
T(p,q)=\frac 1{1+D}\frac{\sin \frac{m+1}2(p-q)}{\sin \frac 12(p-q)}\rho (q).
\label{efpker}
\end{equation}
Here $\rho (q)$ is the quasi-particle Fermi-like distribution function (\ref
{disph}) 
\begin{equation}
\rho (p)=\frac 1{2\pi }(1+D)\left( 1+\exp (\beta \epsilon (p))\right) ^{-1},
\label{Fermi}
\end{equation}
where $\beta ^{-1}$ is the temperature, the total density $D$ is defined by (%
\ref{denph}) and the energy $\epsilon (p)$ is the solution of the non-linear
integral equation (\ref{eenph}). After symmetrizing the kernel the
representation for the darkness formation probability can be rewritten as
follows 
\begin{equation}
\tau (m,\beta )=(1+D)\mbox{det}(\hat I-\hat M),  \label{efpT}
\end{equation}
where $\hat M$ is an integral operator with the kernel 
\begin{equation}
M(p,q)=\frac 1{2\pi }\sqrt{\nu (p,\beta )}\frac{\sin \frac{m+1}2(p-q)}{\sin
\frac 12(p-q)}\sqrt{\nu (q,\beta )},  \label{kernT}
\end{equation}
and  $\nu (p,\beta )=\left( 1+\exp (\beta \epsilon (p))\right) ^{-1}$ is the
Fermi weight.

At zero temperature ($\beta ^{-1}=0$) the Fermi weight becomes the step
function equal to zero outside the Fermi zone and equal to one inside it. 
Thus one has 
\begin{equation}
\tau _0(m)=(1+D_0)\mbox{det}(\hat I-\hat M_0),  \label{efpT0}
\end{equation}
where $\hat M_0$ is an integral operator 
\begin{equation}
(\hat M_0f)(p)=\int\limits_{-\Lambda }^\Lambda M_0(p,q)f(q)dq,
\end{equation}
with the kernel 
\begin{equation}
M_0(p,q)=\frac 1{2\pi }\frac{\sin \frac{m+1}2(p-q)}{\sin \frac 12(p-q)},
\label{kernT0}
\end{equation}
and  the Fermi momentum $\Lambda $ is defined by the equations (\ref{Lambda}).

Consider now the two-point time-dependent correlation function (\ref{corr}%
) in the thermodynamic limit. The equation (\ref{cor1}) takes the form 
\begin{equation}
f^{(\pm )}(m,t,\beta )=\exp \left( \pm \frac D{1+D}\right) \frac{1+D}{2\pi }%
\sum\limits_{l=0}^\infty \int\limits_{-\pi }^\pi e^{i(m-l)\Theta }h^{(\pm
)}(l,t,\beta ,\Theta )d\Theta .  \label{cortherm}
\end{equation}
The functions $h^{(\pm )}(l,t,\beta ,\Theta )$ can be written as Fredholm
determinants 
\begin{equation}
h^{(+)}(l,t,\beta ,\Theta )=\left( G(l,t)+\frac \partial {\partial x}\right) %
\mbox{det}(\hat I+\hat V-x\hat R^{+})|_{x=0},  \label{htherm}
\end{equation}
\begin{equation}
h^{(-)}(l,t,\beta ,\Theta )=\frac \partial {\partial x}\mbox{det}(\hat
I+\hat V+x\hat R^{-})|_{x=0},  \label{htherm-}
\end{equation}
where $\hat V$ and $\hat R^{\pm }$ are integral operators 
\[
(\hat Vf)(p)=\frac 1{2\pi }\int\limits_{-\pi }^\pi V(p,q)f(q)dq,
\]
\begin{equation}
(\hat R^{\pm }f)(p)=\frac 1{2\pi }\int\limits_{-\pi }^\pi R^{\pm
}(p,q)f(q)dq,
\end{equation}
with kernels 
\[
V(p,q)=\frac 1{\tan \frac 12(p-q)}\left( \frac 12(E_{+}^{+}(l,t,p,\beta
,\Theta )+E_{+}^{-}(l,t,p,\beta ,\Theta ))E_{-}(l,t,q,\beta )-\right. 
\]
\[
\left. -\frac 12(E_{+}^{+}(l,t,q,\beta ,\Theta )+E_{+}^{-}(l,t,q,\beta
,\Theta ))E_{-}(l,t,p,\beta )\right) -
\]
\[
-G(l,t)E_{-}(l,t,p,\beta )E_{-}(l,t,q,\beta )+\frac i2(E_{+}^{+}(l,t,p,\beta
,\Theta )-E_{+}^{-}(l,t,p,\beta ,\Theta ))E_{-}(l,t,q,\beta )+
\]
\begin{equation}
+\frac i2(E_{+}^{+}(l,t,q,\beta ,\Theta )-E_{+}^{-}(l,t,q,\beta ,\Theta
))E_{-}(l,t,p,\beta ),  \label{VT}
\end{equation}
\begin{equation}
R^{+}(p,q)=E_{+}^{+}(l,t,p,\beta ,\Theta )E_{+}^{-}(l,t,q,\beta ,\Theta ),
\label{RT}
\end{equation}
\begin{equation}
R^{-}(p,q)=E_{-}(l,t,p,\beta )E_{-}(l,t,q,\beta ).  \label{RT-}
\end{equation}
The functions $G(l,t),E_{-}(l,t,p,\beta ),E_{+}^{+}(l,t,p,\beta ,\Theta )$
and $E_{+}^{-}(l,t,p,\beta ,\Theta )$ are defined as follows 
\begin{equation}
G(l,t)=\frac 1{2\pi }\int\limits_{-\pi }^\pi \exp (ilq-it\varepsilon (q))dq,
\label{GT}
\end{equation}
\begin{equation}
E(l,t,p,\Theta )=\frac 1{2\pi }\mbox{v.p.}\int\limits_{-\pi }^\pi \frac{\exp
(ilq-it\varepsilon (q))}{\tan \frac 12(q-p)}dq+\tan \frac 12(p-\Theta )\exp
(ilp-it\varepsilon (p)),  \label{ET}
\end{equation}
\begin{equation}
E_{-}(l,t,p,\beta )=\sqrt{\nu (p,\beta )}\exp \left( (-i\frac l2p+i\frac
t2\varepsilon (p)\right) ),  \label{E-T}
\end{equation}
\[
E_{+}^{+}(l,t,p,\beta ,\Theta )=\frac 12E_{-}(l,t,p,\beta )\times 
\]
\begin{equation}
\times \left( (E(l,t,p,\Theta )+e^{-i\Theta }E(l+1,t,p,\Theta
))-i(G(l,t)-e^{-i\Theta }G(l+1,t))\right) ,  \label{E++T}
\end{equation}
\[
E_{+}^{-}(l,t,p,\beta ,\Theta )=\frac 12E_{-}(l,t,p,\beta )\times 
\]
\begin{equation}
\times \left( (E(l,t,p,\Theta )+e^{i\Theta }E(l-1,t,p,\Theta
))+i(G(l,t)-e^{i\Theta }G(l-1,t))\right) .  \label{E+-T}
\end{equation}
One should note that the function $E(l,t,p,\Theta )$ is singular at the
points $p=\Theta \pm \pi $ but the functions $E_{+}^{+}(l,t,p,\beta ,\Theta )
$ and $E_{+}^{-}(l,t,p,\beta ,\Theta )$ entering the kernels have no
singularities being well defined for all $\Theta $ and $p$.

In the case of equal-time correlators these functions can be calculated
explicitly 
\begin{equation}
G(l,0)=\delta _{l,0}  \label{GT0}
\end{equation}
\begin{equation}
E_{-}(l,0,p,\beta )=\sqrt{\nu (p,\beta )}e^{-i\frac l2p},  \label{E-T0}
\end{equation}
\begin{equation}
E_{+}^{+}(l,0,p,\beta ,\Theta )=i(1-\delta _{l,0})\sqrt{\nu (p,\beta )}%
e^{i\frac l2p},  \label{E++T0}
\end{equation}
\begin{equation}
E_{+}^{-}(l,0,p,\beta ,\Theta )=i(1-\delta _{l,1})e^{i(\Theta -p)}\sqrt{\nu
(p,\beta )}e^{i\frac l2p},  \label{E+-T0}
\end{equation}
and the kernels $V(p,q)$ and $R(p,q)$ are polynomials in $e^{i\Theta }$.
Hence, $h^{(\pm)}(l,0,\beta ,\Theta )$ can be represented as Taylor series in $%
e^{i\Theta }$ and 
\begin{equation}
\int\limits_{-\pi }^\pi e^{i(m-l)\Theta }h^{(\pm)}(l,0,\beta ,\Theta )d\Theta
=0\makebox[2em]{}{\rm for}\makebox[1em]{}m>l,
\end{equation}
so that
\begin{equation}
f^{(\pm )}(m,0,\beta )=\exp \left( \pm \frac D{1+D}\right) \frac{1+D}{2\pi }%
\sum\limits_{l=m}^\infty \int\limits_{-\pi }^\pi e^{i(m-l)\Theta
}h^{(\pm)}(l,0,\beta ,\Theta )d\Theta .
\end{equation}
The functions $h^{(\pm )}(l,0,\beta ,\Theta )$ can be represented as
determinants of simple operators: 
\begin{equation}
h^{(\pm )}(l,0,\beta ,\Theta )=\frac \partial {\partial x}\mbox{det}(\hat
I-\hat v+x\hat r^{\pm })|_{x=0},  \label{ht=0}
\end{equation}
where the kernels of the operators $\hat v$  and $\hat r^{\pm }$ are 
\begin{equation}
v(p,q)=\sqrt{\nu (p)}\frac{\sin \frac{l+1}2(p-q)+\exp [i(\Theta -\frac{p+q}%
2)]\sin \frac{l-2}2(p-q)}{\sin \frac 12(p-q)}\sqrt{\nu (q)},  \label{vt=0}
\end{equation}
\begin{equation}
r^{+}(p,q)=\sqrt{\nu (p)}e^{i(\Theta -q)}e^{i\frac l2(p+q)}\sqrt{\nu (q)},
\label{rt=0}
\end{equation}
\begin{equation}
r^{-}(p,q)=\sqrt{\nu (p)}e^{-i\frac l2(p+q)}\sqrt{\nu (q)}.  \label{rt=0-}
\end{equation}
Again at zero temperature the Fermi weight  $\nu (p,\beta )$  becomes the
step function and the correlators have the following form 
\begin{equation}
f_0^{(\pm )}(m,0)=\exp \left( \pm \frac{D_0}{1+D_0}\right) \frac{1+D_0}{2\pi 
}\sum\limits_{l=0}^\infty \int\limits_{-\pi }^\pi e^{i(m-l)\Theta }h_0^{(\pm
)}(l,0,\Theta )d\Theta ,
\end{equation}
\begin{equation}
h_0^{(+)}(l,t,\Theta )=\left( G(l,t)+\frac \partial {\partial x}\right) %
\mbox{det}(\hat I+\hat V_0-x\hat R_0^{+})|_{x=0},  \label{h0therm}
\end{equation}
\begin{equation}
h_0^{(-)}(l,t,\Theta )=\frac \partial {\partial x}\mbox{det}(\hat I+\hat
V_0+x\hat R_0^{-})|_{x=0},  \label{h0therm-}
\end{equation}
where $\hat V_0$ and $\hat R_0^{\pm }$ are integral operators 
\[
(\hat V_0f)(p)=\frac 1{2\pi }\int\limits_{-\Lambda }^\Lambda V_0(p,q)f(q)dq,
\]
\begin{equation}
(\hat R_0^{\pm }f)(p)=\frac 1{2\pi }\int\limits_{-\Lambda }^\Lambda R_0^{\pm
}(p,q)f(q)dq.
\end{equation}
The kernels of these integral operators acting on the interval $[-\Lambda
,\Lambda ]$ are given by the equations (\ref{VT}-\ref{RT-}) after putting
formally $\nu (p,\beta )=1$.

\section*{Conclusion}

In this paper we have represented the correlation functions for the phase
model as determinants. In the thermodynamic limit these are the Fredholm
determinants of ''integrable integral operators'' \cite{iiks1}. These
representations will allow us to derive classical integrable equations for
the correlators and to evaluate their large time and
distance asymptotics.

The model considered is not a free fermion model but the correlation
functions are represented in an explicit form not involving auxiliary dual
fields. It is interesting to mention that the corresponding point exists
also for the XXZ Heisenberg chain with an infinite anisotropy. It is natural
that our technique can be applied also at this point, and the expressions
for the correlators will be considerably simpler than in the general case.
We are going to present this results somewhere.

\section*{Acknowledgments}

This work was partially supported by grants No. 95-01-00476a of the Russian
Foundation of Fundamental Research and INTAS-RFBR 95-0414.

\section{Appendix: Form factors}

\renewcommand{\theequation}{A.\arabic{equation}} \setcounter{equation}{0} In
this Appendix we calculate the matrix element 
\begin{equation}
G(\{\lambda _j^B\},\{\lambda _k^C\})=\langle 0|\prod\limits_{k=1}^{N-1}{\bf C%
}(\lambda _k^C)\phi _M\prod\limits_{j=1}^N{\bf B}(\lambda _j^B)|0\rangle .
\label{ff1}
\end{equation}
Using (\ref{split}) with $m=M-1$ and taking into account that in this case $%
B_2=\phi _M^{+},C_2=\phi _M$ we obtain the representation 
\[
G(\{\lambda _j^B\},\{\lambda _k^C\})=\sum\limits_{{\rm I},{\rm II}%
}\prod\limits_{{\rm I}}r_2(\lambda _I^B)_1\langle 0|\prod\limits_{{\rm I}}%
{\bf C_1}(\lambda _I^C)\prod\limits_{{\rm I}}{\bf B_1}(\lambda
_I^B)|0\rangle _1\times 
\]
\begin{equation}
\prod\limits_{{\rm II}}r_1(\lambda _{{\rm II}}^C)\prod\limits_{{\rm II}%
}\frac 1{d_2(\lambda _{{\rm II}}^B)}\prod\limits_{{\rm II}}\frac
1{d_2(\lambda _{{\rm II}}^C)}\left( \prod\limits_{{\rm I}}\prod\limits_{{\rm %
II}}\frac{e^{2\lambda _I^B}}{e^{2\lambda _I^B}-e^{2\lambda _{{\rm II}}^B}}%
\right) \left( \prod\limits_{{\rm I}}\prod\limits_{{\rm II}}\frac{%
e^{2\lambda _{{\rm II}}^C}}{e^{2\lambda _{{\rm II}}^C}-e^{2\lambda _I^C}}%
\right) ,  \label{ff2}
\end{equation}
where the sum is taken over all the partitions of the sets $\{\lambda ^B\}$
and $\{\lambda ^C\}$ into two subsets $\{\lambda _{{\rm I}}^B\}$, $\{\lambda
_{{\rm II}}^B\}$ and $\{\lambda _{{\rm I}}^C\}$, $\{\lambda _{{\rm II}}^C\}$%
, respectively, satisfying the following conditions 
\[
\mbox{card}\{\lambda _{{\rm I}}^B\}=\mbox{card}\{\lambda _{{\rm I}}^C\}=n_1,%
\makebox[2em]{}n_1=0,1,\dots ,N-1, 
\]
\[
\mbox{card}\{\lambda _{{\rm II}}^B\}=\mbox{card}\{\lambda _{{\rm II}%
}^C\}+1=N-n_1\equiv n_2. 
\]
Using (\ref{scprod}) we can rewrite (\ref{ff2}) 
\[
G(\{\lambda _j^B\},\{\lambda _k^C\})=\left( \prod\limits_{j<l}(e^{2\lambda
_j^B}-e^{2\lambda _l^B})\prod\limits_{k<m}(e^{2\lambda _m^C}-e^{2\lambda
_k^C})\right) ^{-1}\times 
\]
\[
\exp (\sum\limits_{j=1}^N\lambda _j^B+\sum\limits_{k=1}^{N-1}\lambda
_k^C)\sum\limits_{{\rm I},{\rm II}}(-1)^{([P]+[Q])}\prod\limits_{{\rm II}%
}\exp (2(M+n_1-1)\lambda _{{\rm II}}^C)\times 
\]
\begin{equation}
\prod\limits_{j<l}(e^{2\lambda _{{\rm II}_j}^B}-e^{2\lambda _{{\rm II}%
_l}^B})\prod\limits_{m<k}(e^{2\lambda _{{\rm II}_k}^C}-e^{2\lambda _{{\rm II}%
_m}^C})\mbox{det}_{n_1}H_1(\{\lambda _I^B\},\{\lambda _I^C\}),  \label{ff3}
\end{equation}
where 
\begin{equation}
H_{1_{jk}}=\frac{\exp (2(N+M-1)\lambda _j^B)-\exp (2(N-n_1+1)\lambda
_j^B+2(M+n_1-2)\lambda _k^C)}{e^{2\lambda _j^B}-e^{2\lambda _k^C}}.
\label{h1}
\end{equation}
Let us consider the function in the left hand side of (\ref{ff3}) 
\[
G_2(\{\lambda _{{\rm II}}^B\},\{\lambda _{{\rm II}}^C\})=\prod\limits_{{\rm %
II}}\exp (2(M+n_1-1)\lambda _{{\rm II}}^C)\prod\limits_{j<l}(e^{2\lambda _{%
{\rm II}_j}^B}-e^{2\lambda _{{\rm II}_l}^B})\prod\limits_{m<k}(e^{2\lambda _{%
{\rm II}_k}^C}-e^{2\lambda _{{\rm II}_m}^C}). 
\]
These products can be rewritten as determinants 
\[
G_2(\{\lambda _{{\rm II}}^B\},\{\lambda _{{\rm II}}^C\})=\prod\limits_{{\rm %
II}}\exp (2(M+n_1-1)\lambda _{{\rm II}}^C)\mbox{det}_{N-n_1}h(\{\lambda _{%
{\rm II}}^B\})\mbox{det}_{N-n_1-1}g(\{\lambda _{{\rm II}}^C\}), 
\]
where the matrices $g$ and $h$ are given by the formulae 
\[
h_{jk}=e^{2(k-1)\lambda _j^B},\makebox[2em]{}g_{jk}=e^{2(N-n_1-j-1)\lambda
_k^C}. 
\]
It is convenient to introduce the $(N-n_1)\times (N-n_1)$ matrix $\tilde h$ 
\[
\tilde h_{1k}=\tilde h_{k1}=\delta _{1k},\makebox[1em]{}\makebox[1em]{}%
\tilde h_{jk}=g_{j-1,k-1}\makebox[1em]{}{\rm for}\makebox[1em]{}j>1,k>1, 
\]
\[
\mbox{det}_{N-n_1}\tilde h=\mbox{det}_{N-n_1-1}g. 
\]
Then it is easy to see that 
\[
G_2(\{\lambda _{{\rm II}}^B\},\{\lambda _{{\rm II}}^C\})=\mbox{det}%
_{N-n_1}H_2(\{\lambda _{{\rm II}}^B\},\{\lambda _{{\rm II}}^C\}), 
\]
\[
H_{2_{j1}}=1, 
\]
\begin{equation}
H_{2_{jk}}=\frac{\exp (2(n_1+M-1)\lambda _{k-1}^C+2(N-n_1)\lambda _j^B)-\exp
(2\lambda _j^B+2(M+N-2)\lambda _{k-1}^C)}{e^{2\lambda _j^B}-e^{2\lambda
_{k-1}^C}},  \label{h2}
\end{equation}
for $k>1$. The representation (\ref{ff3}) can be rewritten now in the
following form 
\[
G(\{\lambda _j^B\},\{\lambda _k^C\})=\left( \prod\limits_{j<l}(e^{2\lambda
_j^B}-e^{2\lambda _l^B})\prod\limits_{k<m}(e^{2\lambda _m^C}-e^{2\lambda
_k^C})\right) ^{-1}\times 
\]
\[
\exp (\sum\limits_{j=1}^N\lambda _j^B+\sum\limits_{k=1}^{N-1}\lambda
_k^C)\sum\limits_{{\rm I},{\rm II}}(-1)^{([P]+[Q])}\mbox{det}%
_{n_1}H_1(\{\lambda _I^B\},\{\lambda _I^C\})\times 
\]
\begin{equation}
\times \mbox{det}_{N-n_1}H_2(\{\lambda _{{\rm II}}^B\},\{\lambda _{{\rm II}%
}^C\}).  \label{ff4}
\end{equation}
To use the Laplace formula for the determinant of the sum of two matrices it
is necessary to introduce four dual fields \cite{kor}: $\psi _j^{+},\psi _j,%
\makebox[1em]{}j=1,2,3,4$ 
\[
\lbrack \psi _j,\psi _k]=[\psi _j^{+},\psi _k^{+}]=0,\makebox[1em]{}[\psi
_j,\psi _k^{+}]=\delta _{j,k}, 
\]
acting in the dual Fock space with the vacuum $|0)$ 
\[
\psi _j|0)=0,\makebox[1em]{}(0|\psi _j^{+}=0. 
\]
Using the commutation relation for the dual fields it is easy to prove the
following relations 
\[
\mbox{det}_{n_1}H_1(\{\lambda _I^B\},\{\lambda _I^C\})=(0|\mbox{det}_{n_1}%
{\cal G}_1(\{\lambda _I^B\},\{\lambda _I^C\})|0), 
\]
\begin{equation}
\mbox{det}_{N-n_1}H_2(\{\lambda _{{\rm II}}^B\},\{\lambda _{{\rm II}%
}^C\})=(0|\mbox{det}_{N-n_1}{\cal G}_2(\{\lambda _{{\rm II}}^B\},\{\lambda _{%
{\rm II}}^C\})|0),  \label{detfield2}
\end{equation}
\[
\label{g1}{\cal G}_{1_{jk}}=\frac{\exp (\psi _2^{+}+\psi _1)}{e^{2\lambda
_j^B}-e^{2\lambda _k^C}}(\exp (2(N+M-1)\lambda _j^B)- 
\]
\begin{equation}
-\exp (2(N+1)\lambda _j^B+2(M-2)\lambda _k^C+2(\lambda _k^C-\lambda
_j^B)(\psi _1^{+}+\psi _2)),
\end{equation}
\[
{\cal G}_{2_{j1}}=1, 
\]
\[
{\cal G}_{2_{jk}}=\frac{\exp (\psi _4^{+}+\psi _3)}{e^{2\lambda
_j^B}-e^{2\lambda _{k-1}^C}}\exp (2(N+M-1)\lambda _{k-1}^C+2(\lambda
_j^B-\lambda _{k-1}^C)(\psi _3^{+}+\psi _4))- 
\]
\begin{equation}
-\exp (2\lambda _j^B+2(M+N-2)\lambda _{k-1}^C),\makebox[1em]{}{\rm for}%
\makebox[1em]{}k>1.  \label{g2}
\end{equation}
Now (\ref{ff4}) can be rewritten in the form
\[
G(\{\lambda _j^B\},\{\lambda _k^C\})=\left( \prod\limits_{j<l}(e^{2\lambda
_j^B}-e^{2\lambda _l^B})\prod\limits_{k<m}(e^{2\lambda _m^C}-e^{2\lambda
_k^C})\right) ^{-1}\times 
\]
\begin{equation}
\times \exp (\sum\limits_{j=1}^N\lambda _j^B+\sum\limits_{k=1}^{N-1}\lambda
_k^C)(0|\mbox{det}_N{\cal G}(\{\lambda ^B\},\{\lambda ^C\})|0),  \label{ff5}
\end{equation}
\[
{\cal G}_{j1}=1,\makebox[1em]{}{\cal G}_{jk}={\cal G}_{1_{j,k-1}}+{\cal G}%
_{2_{jk}}\makebox[1em]{}{\rm for}\makebox[1em]{}k>1. 
\]
The determinant in (\ref{ff5}) is represented as a sum of minors 
\[
(0|\mbox{det}_N{\cal G}(\{\lambda ^B\},\{\lambda
^C\})|0)=\sum\limits_{l=1}^N\exp (\sum\limits_{j\neq l}\lambda
_j^B-\sum\limits_{k=1}^{N-1}\lambda _k^C)\times 
\]
\begin{equation}
\times (-1)^l(0|\mbox{det}_{N-1}V(\{\lambda _1^B,\dots \lambda
_{l-1}^B,\lambda _{l+1}^B,\dots ,\lambda _N^B\}\{\lambda ^C\})|0).
\label{ff6}
\end{equation}
The mean value of the determinant of the matrix $V$ can be expressed without
dual fields, 
\begin{equation}
(0|\mbox{det}_NV(\{\lambda _j^B\},\{\lambda _k^C\})|0)=\mbox{det}%
_NF(\{\lambda _j^B\},\{\lambda _k^C\}).  \label{nofield}
\end{equation}
Now we can use this relation to rewrite $G(\{\lambda _j^B\},\{\lambda
_k^C\}) $ also without dual fields: 
\[
(0|\mbox{det}_N{\cal G}(\{\lambda ^B\},\{\lambda ^C\})|0)=\mbox{det}%
_NH(\{\lambda _j^B\},\{\lambda _k^C\}), 
\]
\[
H_{j1}=1, 
\]
\begin{equation}
H_{jk}=\frac{\exp (2(N+M-1)\lambda _j^B)-\exp (2\lambda _j^B+2(M+N-2)\lambda
_{k-1}^C)}{e^{2\lambda _j^B}-e^{2\lambda _{k-1}^C}}\makebox[1em]{}{\rm for}%
\makebox[1em]{}k>1.
\end{equation}
Finally we have 
\[
G(\{\lambda _j^B\},\{\lambda _k^C\})=\left( \prod\limits_{j<l}(e^{2\lambda
_j^B}-e^{2\lambda _l^B})\prod\limits_{k<m}(e^{2\lambda _m^C}-e^{2\lambda
_k^C})\right) ^{-1}\times 
\]
\begin{equation}
\times \exp (\sum\limits_{j=1}^N\lambda _j^B+\sum\limits_{k=1}^{N-1}\lambda
_k^C)\mbox{det}_NH(\{\lambda _j^B\},\{\lambda _k^C\}).  \label{ff7}
\end{equation}
If $\{\lambda _j^B\},\{\lambda _k^C\}$ are solutions of the Bethe equations (%
\ref{betheph}) the matrix $H$ can be rewritten as
\begin{equation}
H_{jk}=-\frac{e^{i(P-p_j)}+e^{i(Q-q_{k-1}+p_j)}}{e^{ip_j}-e^{iq_{k-1}}},
\label{h}
\end{equation}
where 
\[
p_j=-2i\lambda _j^B,\makebox[1em]{}q_k=-2i\lambda _k^C,\makebox[1em]{}%
P=\sum\limits_{j=1}^Np_j,\makebox[1em]{}Q=\sum\limits_{k=1}^{N-1}q_k. 
\]
Using (\ref{gaudin}) we obtain the normalized form factor 
\begin{equation}
G_N(M,\{p\},\{q\})=Z_MM^{-1}(M+N)^{-N+\frac 32}\left( \frac{M+N}{M+N-1}%
\right) ^{\frac N2-1}\mbox{det}H(\{p\},\{q\}),  \label{Mff}
\end{equation}
where $Z_M$ is a complex number depending on $P,Q$ and $M$, $|Z_M|=1$.

If $\{\lambda _j^B\},\{\lambda _k^C\}$ are solutions of the Bethe equation
then, using the shift operator, it is also possible to calculate the
normalized matrix elements of any operator $\phi _m$. 
\begin{equation}
G_N(m,\{p\},\{q\})=Z_mM^{-1}(M+N)^{-N+\frac 32}\left( \frac{M+N}{M+N-1}%
\right) ^{\frac N2-1}\mbox{det}H(\{p\},\{q\}),  \label{mff}
\end{equation}
\[
Z_m=e^{i(M-m)(Q-P)}Z_M. 
\]
Notice that this representation is equivalent to the formulae (\ref{difff}),(%
\ref{d1d2}). It is evident that now one can also obtain a representation for
the normalized matrix elements of the operators $\phi _m^{\dagger }$, using
the following relation 
\[
G_N^{(+)}(m,\{p\},\{q\})=G_{N+1}^{*}(m,\{q\},\{p\}), 
\]
where star means the complex conjugation. Thus one gets for this form
factors the representation (\ref{ff+}),(\ref{d1+d2+}).

\end{document}